\documentclass{emulateapj}
\usepackage{amsmath,natbib,psfig}

\input{buote.defs}

\begin{document} 

\title{The Heated Core of the Radio-Quiet Galaxy Cluster A644}
\author{David A. Buote\altaffilmark{1}, Philip
J. Humphrey\altaffilmark{1}, \& John T. Stocke\altaffilmark{2}}
\altaffiltext{1}{Department of Physics and Astronomy, University of
California at Irvine, 4129 Frederick Reines Hall, Irvine, CA 92697-4575}
\altaffiltext{2}{Center for Astrophysics and Space Astronomy,
University of Colorado, 389 UCB,  Boulder, CO 80309}

\slugcomment{Accepted for Publication in The Astrophysical Journal}

\begin{abstract}
We present an analysis of a \chandra\ ACIS-I observation of the
massive galaxy cluster A644. This cluster was previously classified as
a cooling flow, but no radio emission has been detected from its cD
galaxy. Outside the core ($R\sim 75$~kpc $\sim 0.03\rvir$) the hot ICM
has properties consistent with a (relaxed) cool-core cluster out to
the largest radii investigated ($R\sim 415$~kpc $\sim
0.14\rvir$). Over this region the gravitating mass profile is
described well by a Navarro-Frenk-White profile with concentration
parameter, $c=6.1\pm 1.2$, and virial radius, $\rvir=2.9\pm 0.4$~Mpc.
However, inside the core the temperature and entropy profiles reverse
their inward radial decline and rise at the center; the inner
temperature profile is inconsistent with a constant at the $2.3\sigma$
level. Although the core region does not display X-ray cavities or
filamentary structures characteristic of radio-loud, cool-core
clusters, the peak of the X-ray emission is offset from that of the
centroid of the global X-ray halo by $\approx 60$~kpc. The position of
the cD galaxy lies approximately between the X-ray peak and centroid,
further testifying to a merger origin for the properties of the X-ray
emission in the core. We discuss the implications of A644 and the
small number of radio-quiet, cool-core clusters for the AGN feedback
paradigm to suppress cooling flows in clusters.
\end{abstract}

\keywords{X-rays: galaxies: clusters -- galaxies: halos -- galaxies:
formation -- cooling flows -- galaxies: clusters: individual: A644} 

\section{Introduction}
\label{intro}

\xmm\ and \chandra\ grating and CCD observations of the cores of clusters
previously believed to be harboring cooling flows find very little gas
cooling below $T\sim 1$~keV
\citep[e.g.,][]{pete01,pete03a,tamu01,davi01a,mole01b,xu02a,etto02a,buot03a}.
This major discovery has led to the widely discussed picture where
feedback on the hot intracluster medium (ICM) from an AGN in the
central cluster galaxy is responsible for stifling the cooling flow,
though the details of the heating process remain elusive \citep[e.g.,
for reviews see][]{math03a,fabi04a}. About 70\% of clusters previously
classified as cooling flows possess radio emission from the central
galaxy \citep[e.g.,][]{burn90a}. \chandra\ observations indicate that
such clusters also have X-ray cavities and filamentary structures in
the X-ray images that are related to the radio emission
\citep[e.g.,][]{fabi00_perseus,mcna00a,blan01,mazz03a,birz04a}. 

What about clusters indicated to be cooling flows from previous
low-resolution X-ray observations that do not display radio properties
typical of cooling flows? We have previously analyzed \chandra\
observations of two such systems: A2029 \citep{lewi02a,lewi03a} and
A2589 \citep{buot04a}. A2589 has no observed radio emission associated
with its cD galaxy; and, although the cD in A2029 does possess a
Wide-Angle-Tail (WAT) radio source, such sources are generally not
found in cooling flows \citep{owen84a,burn90a}, though their
morphology provides evidence for relative motions between the radio
source and ICM \citep{eile84a,burn02a}. Both of these clusters have
high density cores with temperatures significantly lower than at
larger radii. (Although the relatively low-quality \chandra\ data of
A2589 are consistent with an isothermal gas, a new \xmm\ observation
clearly indicates that the temperature profile increases with radius
-- Zappacosta et al.\ 2005, in preparation.)  These clusters also
display no spectroscopic evidence for cooling flows and have unusually
symmetric X-ray images without the pronounced surface brightness
irregularities observed in the cores of ``radio loud'' cool-core
clusters \citep[e.g.,][]{birz04a}. Finally, cool-core clusters also
appear to preferentially display central metallicity enhancements, while
flat metallicity profiles are characteristic of non-cooling flow,
generally disturbed, systems \citep[e.g.,][]{degr04a,bohr04a}. A2589
and A2029 both have pronounced central metallicity enhancements
further attesting to their comparatively relaxed states.

Observations suggest that cooling flows are
destroyed by large turbulent motions in the ICM since major merging
activity in clusters anti-correlates with cooling mass-flow rate
\citep{buot96b}, a conclusion supported by theoretical studies
\citep[e.g.,][]{roet96a,norm99a,fuji04a}. Likewise, theoretical
studies \citep[e.g.,][]{loke95a,roet96a} suggest that WAT radio
morphology may also be a signature of hot shocks in on-going cluster
mergers.  Since the X-ray images of A2029 and A2589 are very regular
all the way into their cores (though see \S \ref{conc}), and yet there
is no evidence for ICM cooling or radio emission typical of cool-core
clusters, in the context of the AGN feedback paradigm we are viewing
these clusters at a special time after they have settled down from
past mergers but very soon before the onset of AGN heating.

Here we examine a \chandra\ observation of A644, another ``radio
quiet'' cluster that was previously classified as a cooling flow based
on analysis of both X-ray imaging and spectral data
\citep{edge92a,pere98a}.  A644 is the only cooling flow cluster with
large inferred $\mdot$ ($>$ 100 $\msun$ yr$^{-1}$), which was not
detected in the radio study of
\citet[][$P_{\rm 6 cm}<$ 3 x10$^{21}$ W Hz$^{-1}$]{burn90a}.  We wish to
examine how its X-ray properties compare to the other cooling flow
clusters with atypical radio properties and, in particular, whether it
demonstrates a clearer connection between past merger activity and the
suppression of a cooling flow.  The redshift of A644 ($z=0.0704$)
corresponds to an angular diameter distance of 277~Mpc and $1\arcsec =
1.34$~kpc assuming $\omegam=0.3$, $\omegalambda=0.7$, and $H_0 =
70$~\kmsmpc. 

The paper is organized as follows. In \S \ref{obs} we present the
observation and discuss the data reduction. The analysis of the image
and spectral data of the ACIS-I are presented in \S \ref{image} and \S
\ref{spec} respectively. We calculate the gravitating matter
distribution in \S \ref{mass}. Finally, in \S \ref{conc} we present
our conclusions.

\section{Observations and Data Analysis}
\label{obs}

A644 was observed with the ACIS-I CCD array for approximately 30~ks
during AO-2 as part of the \chandra\ Guest Observer program. We
reduced the data using the {\sc CIAO} v3.1 (with {\sc CALDB} v2.28)
and {\sc HEASOFT} v5.3.1 software packages. We followed the standard
{\sc CIAO} threads\footnote{http://asc.harvard.edu/ciao/threads} and
reprocessed the level-1 events data to make use of the latest
calibration information, including correcting for charge-transfer
inefficiency and a time-dependent gain shift. From regions of least
source contamination of the CCDs we extracted a light-curve
(5.0-10.0~keV) to identify periods of high background. The observation
was remarkably quiescent, and after removing a small period of
modestly increased background rate, the total exposure used for
subsequent analysis was 28.9~ks.

Since the cluster emission fills the entire ACIS-I field, we adopted
for our default analysis the standard background events files provided
in the {\sc CALDB} collected from suitable blank fields. For
comparison to results obtained using these standard background
templates, we also modeled the background directly. Using regions as
far away from the cluster center as possible, we extracted a spectrum
(with point sources masked out) and fitted it with a combined source
and background model \citep[e.g.,][]{buot04b}. For the source, we
adopted a thin thermal plasma component (\apec) with $T=6$~keV and
half-solar metallicity. The background was modeled with components for
the Cosmic X-ray Background (CXB) and instrumental background: two
soft \apec\ thermal plasmas and a hard power law for the CXB; a broken
power law and two gaussians for the instrumental
background. Comparison of results using the alternative background is
presented in \S \ref{spec}.

\section{Imaging Analysis}
\label{image}

\begin{figure*}[t]
\parbox{0.49\textwidth}{
\centerline{\psfig{figure=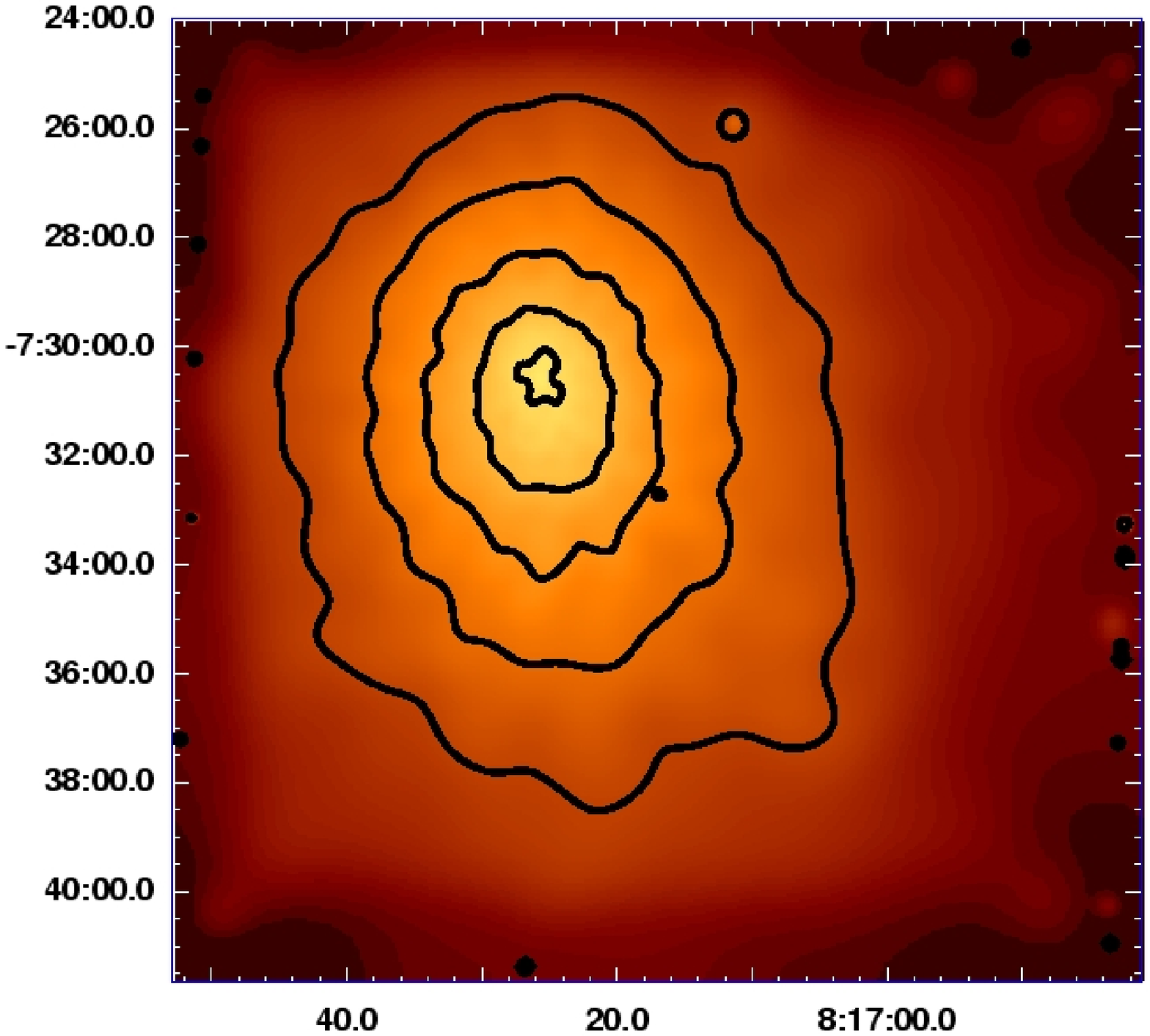,angle=0,height=0.33\textheight}}}
\parbox{0.49\textwidth}{
\centerline{\psfig{figure=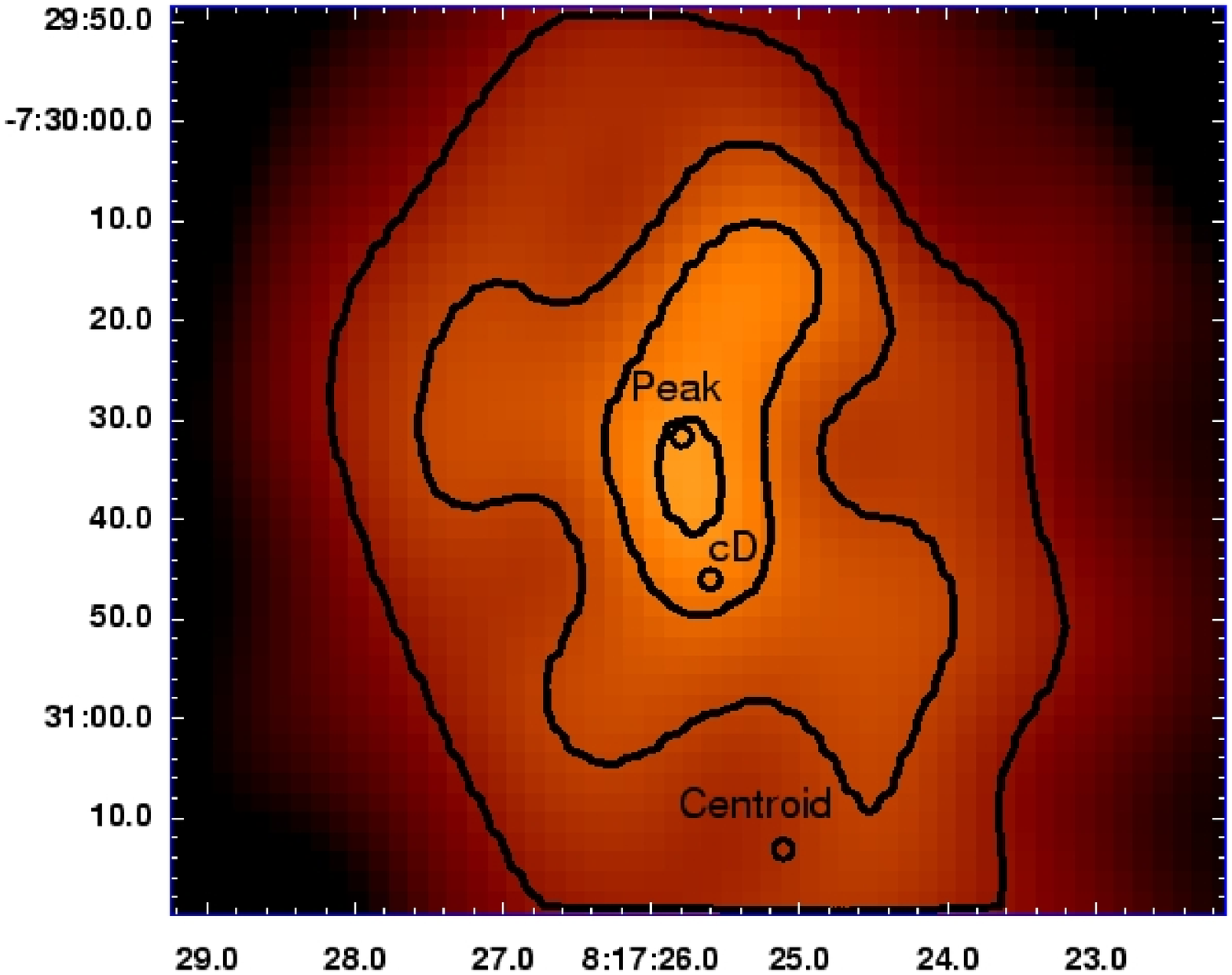,angle=0,height=0.28\textheight}}}
%\vspace{-1.0cm}
\caption{\label{fig.image} \footnotesize
({\it Left}) Smoothed, exposure-corrected ACIS-I image of A644 in the
0.3-7.0~keV band before point sources have been removed. Overlaid
contours are logarithmically spaced in intensity. ({\it Right})
Close-up of the $R\approx 45\arcsec$ region centered about the
emission peak. The small circles indicate the (J2000) positions of the
peak (08h17m25.793s,-07d30m31.69s) and centroid
(08h17m25.106s,-07d31m13.14s) determined from the X-ray data and the
optical position of the cD galaxy (08h17m25.6s,-07d30m46s). Note that
the peak is computed by centroiding on the 30 brightest pixels.}
%\vspace{-0.5cm}
\end{figure*}

In Figure \ref{fig.image} we display the ACIS-I image of A644 in the
0.3-7.0~keV band. The image was corrected for exposure variations
using an exposure map created for a monochromatic energy 1~keV,
approximately the counts-weighted average energy of the spectrum. For
presentation purposes only, the image was also adaptively smoothed
using the {\sc CIAO} task {\sc csmooth} with default parameter
settings. The diffuse emission is seen to fill the entire ACIS-I
field. The morphology of most of the cluster is quite regular and
approximately elliptical in shape.

Visual inspection of the central $R\sim 30\arcsec$ region of the
(smoothed) X-ray surface brightness reveals some irregular,
non-azimuthally symmetric features (Figure \ref{fig.image}). These
deviations from a symmetric configuration, though statistically
significant, do not represent large fluctuations ($<10\%$) in the mean
surface brightness level. In particular, the data do not provide clear
evidence for distinct subclumps in the core region.

We also searched for azimuthal fluctuations in the core by
constructing a hardness ratio map (and also through spectral fitting
in sectors, see \S \ref{spec}). We did not find any corresponding
azimuthal fluctuations either in the core or on larger
scales. Instead, we found the spectrally softer core gives way at
larger radius to a spectrally harder surrounding medium. This radial
variation is consistent with the radially increasing temperature
profile (see \S \ref{spec}).

We measure the centroid and the ellipticity of the X-ray surface
brightness using the moment method described by \citet{cm} and
implemented in our previous X-ray studies of galaxies and clusters
\citep[e.g.,][]{buot94}. This iterative method is equivalent to computing the
(two-dimensional) principal moments of inertia within an elliptical
region. The ellipticity is defined by the square root of the ratio of
the principal moments, and the position angle is defined by the
orientation of the larger principal moment. Following our previous
study of the ellipticity of the \chandra\ data of NGC 720
\citep{buot02b} we removed point sources and replaced them with
smoothly distributed diffuse emission using the {\sc CIAO} task {\sc
dmfilth}.

The centroid of the X-ray surface brightness shifts significantly
(e.g., $\Delta R=56\pm 6$~kpc between 15-500~kpc), mostly directed
toward the South. At the smallest radii the centroid is located within
the region of image irregularities in the core mentioned above.
Outside of the core region the X-ray isophotes are very regular and
moderately elliptical.  The ellipticity gently falls from $0.30 \pm
0.02$ at $a=100$~kpc to $0.243 \pm 0.004$ at $a=500$~kpc, where $a$ is
the semi-major axis of the elliptical aperture within which the
moments are calculated. Over this radius range the position angle
(measured N through E) of the elliptical apertures is quite steady,
varying between a maximum of $11\pm 2$ degrees (at $a=100$~kpc) and a
minimum of $6\pm 0.5$ degrees (at $a=300$ kpc). Overall, the image
properties suggest a relaxed cluster outside about $R\approx 100$~kpc
but a disturbed cluster within $R\approx 50$~kpc.

We examined the radial surface brightness profile (with point sources
masked out) located about the X-ray peak and also about the centroid
computed within $a=500$~kpc. In both cases the surface brightness
profile is fitted fairly well out to $R=600$~kpc by a $\beta$ model:
$r_c = 79\pm 2$~kpc, $\beta = 0.50\pm 0.01$, $\chi^2=200.4$, 129 dof
(peak) and $r_c = 118\pm 3$~kpc, $\beta = 0.59\pm 0.01$,
$\chi^2=198.6$, 128 dof (centroid). The largest fit residuals ($\sim
20\%$) lie within the innermost region ($r\la 10\arcsec$) and are
modest ($<5\%$) elsewhere. We note that the parameters $r_c$ and
$\beta$ increase systematically when we include data from increasingly
larger radii or exclude data from the very central region. Both of
these issues likely account for the different values obtained from
previous \rosat\ studies which included data at larger radii; e.g.,
using the PSPC \citet{etto99a} obtain $r_c=157$~kpc and $\beta=0.69$.

\section{Spectral Analysis}
\label{spec}

\begin{figure*}[t]
\parbox{0.49\textwidth}{
\centerline{\psfig{figure=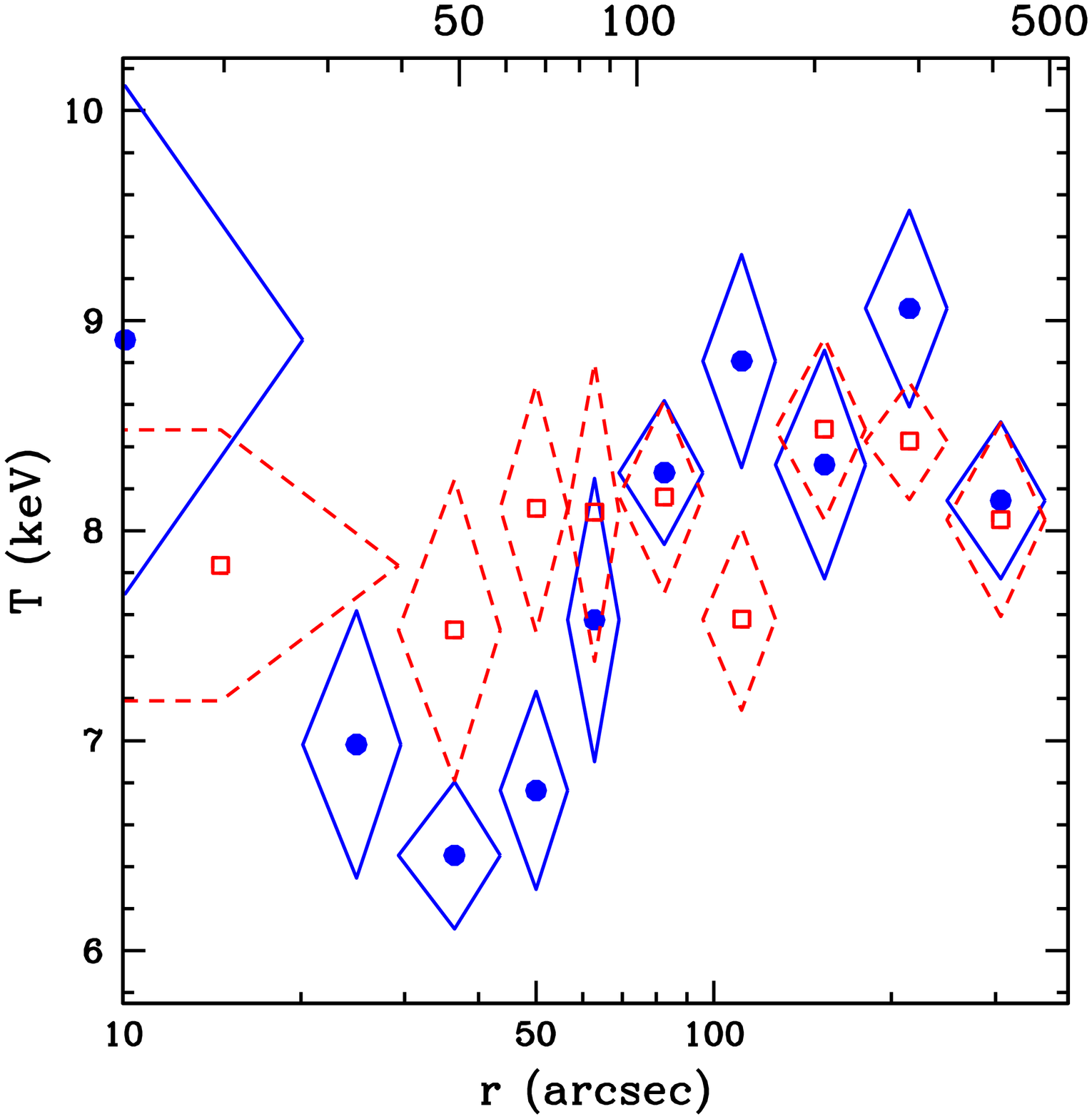,angle=0,height=0.30\textheight}}}
\parbox{0.49\textwidth}{
\centerline{\psfig{figure=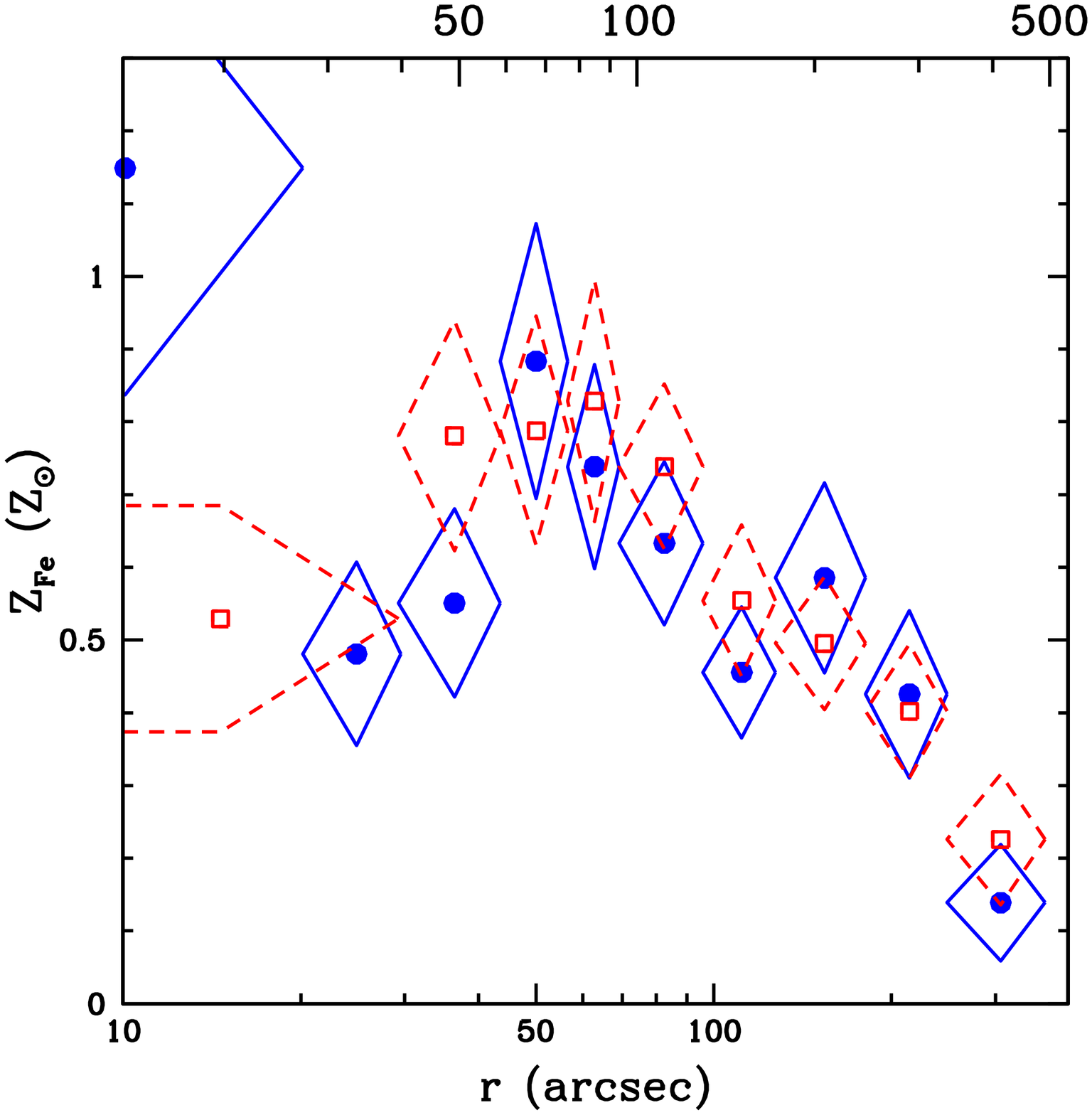,angle=0,height=0.30\textheight}}}
\caption{\label{fig.tfe} \footnotesize
({\it Left}) Radial temperature and ({\it Right}) metallicity profiles
computed within circular annuli oriented about two different center
positions: peak (circles: blue) and centroid (squares: red), as
explained in the text. The top axis indicates radial units in kpc. The
centroid results are also presented in Table \ref{tab.spec}.}
\vspace{-0.5cm}
\end{figure*}

\begin{table*}[t] \footnotesize
\begin{center}
\caption{Parameters from the Spectral Fits}
\label{tab.spec}
\begin{tabular}{c|ll|ll|c|c|c|c}  \tableline\tableline\\[-7pt]
& \multicolumn{2}{c}{$R_{\rm in}$} & \multicolumn{2}{c}{$R_{\rm out}$} & $T$& $\fe$ & $norm$\\
Annulus & (arcsec) & (kpc) & (arcsec) & (kpc) & (keV) & (solar) & (10$^{-3}$~cm$^{-5}$) & $(\chi^2$/dof)\\
\tableline \\[-7pt]
1 & 0.0   &    0.0 & 29.3   & 39.3  & $7.8\pm 0.7$ & $0.53\pm 0.16$ & $2.54\pm 0.07$ & $127.8/140$ \\
2 & 29.3  &   39.3 & 43.5   & 58.5  & $7.5\pm 0.7$ & $0.78\pm 0.16$ & $2.63\pm 0.06$ & $118.3/146$ \\
3 & 43.5  &   58.5 & 56.6   & 76.0  & $8.1\pm 0.6$ & $0.79\pm 0.16$ & $2.86\pm 0.09$ & $159.2/159$ \\
4 & 56.6  &   76.0 & 69.1   & 92.9  & $8.1\pm 0.7$ & $0.83\pm 0.17$ & $2.94\pm 0.10$ & $168.6/165$ \\
5 & 69.1  &   92.9 & 95.9   & 128.9 & $8.2\pm 0.5$ & $0.74\pm 0.12$ & $6.23\pm 0.13$ & $262.7/249$ \\
6 & 95.9  &  128.9 & 127.2  & 170.9 & $7.6\pm 0.4$ & $0.55\pm 0.11$ & $6.01\pm 0.12$ & $235.9/225$ \\
7 & 127.2 &  170.9 & 180.8  & 242.9 & $8.5\pm 0.4$ & $0.45\pm 0.09$ & $7.92\pm 0.16$ & $245.9/265$ \\
8 & 180.8 &  242.9 & 248.2  & 333.5 & $8.4\pm 0.3$ & $0.40\pm 0.09$ & $6.78\pm 0.15$ & $254.4/248$ \\
9 & 248.2 &  333.5 & 364.1  & 489.2 & $8.1\pm 0.5$ & $0.23\pm 0.09$ & $6.65\pm 0.10$ & $217.6/229$ \\
\tableline \\
\end{tabular}
\tablecomments{Results of fitting a single \apec\ plasma emission
model modified by Galactic absorption directly to the annular spectra
(i.e., without spectral deprojection) over 1.5-7~keV for the centroid
case. The $norm$ parameter is the emission measure of the \apec\ model
as defined in
\xspec: $10^{-14}(\int n_en_pdV)/4\pi D^2(1+z)^2$ with units $\rm
cm^{-5}$. The quoted errors are $1\sigma$ computed using the Monte
Carlo procedure described in \S \ref{spec}.  }
\end{center}
\end{table*}

\begin{figure*}[t]
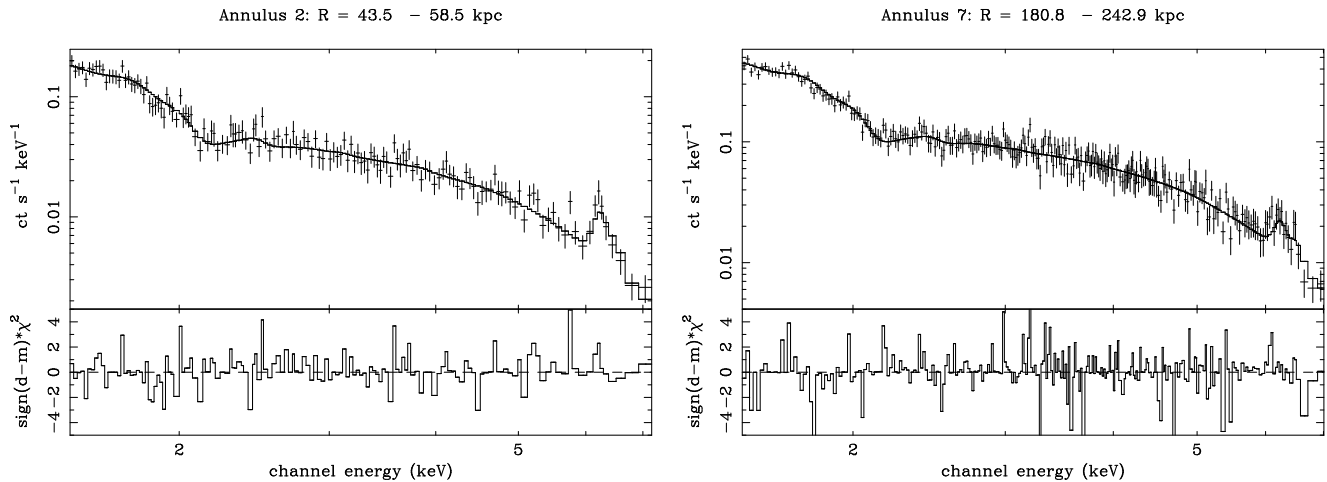

\parbox{0.49\textwidth}{
\centerline{\psfig{figure=f3a.eps,angle=-90,height=0.26\textheight}}}
\parbox{0.49\textwidth}{
\centerline{\psfig{figure=f3b.eps,angle=-90,height=0.26\textheight}}}
\caption{\label{fig.spec} ACIS-I spectra accumulated within
({\sl Left Panel}) annulus \#2 and ({\sl Right Panel}) annulus \#7 for
annuli located about the centroid (Table \ref{tab.spec}). Each
spectrum is fitted with an \apec\ plasma model modified by 
Galactic absorption over 1.5-7~keV as discussed in \S \ref{spec}. }
\end{figure*}

We focus our attention on the spectral properties of the hot gas
obtained within a series of concentric circular annuli centered either
on the X-ray centroid or peak. Below we also briefly discuss attempts
to constrain azimuthal variations in the spectral properties. The
widths of the circular annuli were chosen so that the temperatures
were determined to similar precision in each radial bin. For the
centroid case we list the annuli in Table \ref{tab.spec}. Annuli 1-4
contain $10^4$ background-subtracted counts while bins 5-9 contain
twice that amount in the 1.5-7~keV band. We restrict our analysis to
energies above 1.5~keV because of uncertainties arising from excess
absorption above the Galactic value discussed below, and because the
gas temperature lies well above 1.5~keV within all annuli. For each
annulus counts-weighted response matrices were generated using the
{\sc CIAO} tasks {\sc mkwrmf} and {\sc mkarf}.

We fitted the background-subtracted spectrum with an
\apec\ thermal plasma modified by Galactic absorption ($6.4\times
10^{20}$~\cmsq, \citealt{dick90}) to each annulus. The free parameters
are temperature, iron abundance, and normalization (emission
measure). (The fitted abundance is technically a metallicity since we
fit the iron abundance as a free parameter and the abundances of all
the other elements tied to iron in their solar ratios. However, iron
dominates the metallicity.) The spectral fitting was performed with
\xspec\ \citep[11.3.1i,][]{xspec} using the $\chi^2$ minimization
method.  Hence, we rebinned all spectral channels to have a minimum of
20 counts per energy bin (necessary for the validity of the $\chi^2$
method) and a signal-to-noise ratio of at least 3. The solar
abundances in \xspec\ are taken to be those given by \citet{grsa}
which use the correct ``new'' photospheric value for iron which agrees
also with the value obtained from solar-system meteorites
\citep[e.g.,][]{mcwi97}.  To estimate the statistical errors on the
fitted parameters we simulated spectra for each annulus using the
best-fitting models and fit the simulated spectra in exactly the same
manner as done for the actual data. From 100 Monte Carlo simulations
we compute the standard deviation for each free parameter which we
quote as the ``$1\sigma$'' error. (We note that these $1\sigma$ error
estimates generally agree very well with those obtained using the
standard $\Delta\chi^2$ approach in \xspec.) All quoted errors are
$1\sigma$ unless stated otherwise.

The radial profiles of temperature and iron abundance are shown in
Figure \ref{fig.tfe}; Table \ref{tab.spec} gives the parameters
obtained from the spectral fits for the centroid case; Figure
\ref{fig.spec} shows the ACIS-I spectra of two representative annuli
and the associated best-fitting one-temperature model. The simple
one-temperature model provides a good description of the spectral data
in all annuli. For $R\ga 75$~kpc the spectral fits for both the
centroid and peak case agree very well. Over this region the
temperature is consistent with a constant value $\sim 8$~keV as found
also by previous studies of A644 with \asca\
\citep[e.g.,][]{daw00,baue00b}. In contrast to these previous
\asca\ studies, the \chandra\ data clearly reveal that the iron abundance
declines with radius from a value of $\sim 0.8\solar$ at $R\sim
75$~kpc to $\sim 0.25\solar$ at $R\sim 400$~kpc. Previous \asca\
studies of A644 were consistent with a constant iron abundance profile
\citep[e.g.,][]{daw00,baue00b} within the large errors. We note also that the
average iron abundance from \asca\ ($\approx 0.5\solar$, scaled to our
solar reference) obtained by \citet{daw00} agrees well with our
\chandra\ results. 

For $R\ga 75$~kpc the declining radial profile of iron abundance,
approximately isothermal gas profile, and regular image morphology are
very characteristic of relaxed ``cool core'' clusters
\citep[e.g.,][]{mole01b,degr04a}. Inspection of Figure \ref{fig.tfe}
reveals that the region $R\la 75$~kpc is more complex. The iron
abundance profile is approximately constant for the centroid case,
with marginal evidence for a central dip. In the peak case, the
profile declines and then appears to increase in the central bin: the
value in the central bin ($1.15\pm 0.31$ solar) is inconsistent with
that in the adjacent bin at the $\approx 2.5\sigma$ level.

The temperature profile for $R\la 75$~kpc does not correspond to a
typical cool-core cluster. In the centroid case, the temperature is
consistent with isothermal. In the peak case, the temperature declines
inward to $R\sim 50$~kpc like a cool-core cluster but then rises
significantly in the central bin ($T=8.9\pm 1.2$~keV). The 99\%
confidence lower limit on the temperature estimated using
$\Delta\chi^2=6.63$ is $T=6.7$~keV, still larger than the minimum
value of $T=6.5$~keV reached near $R=50$~kpc. A power law fit to the
temperature profile of the inner three radial bins is inconsistent
with a constant value at the $2.3\sigma$ level and is highly
inconsistent with the declining temperature central profiles
characteristic of cool core clusters.

Despite the good quality of the fits, we examined whether they could
be improved further. Allowing abundances other than iron to vary
separately did not yield noticeable improvement. Adding another
temperature component also had no effect on the fits, even when
energies down to 0.5~keV were included. Consequently, we found no need
for an additional multiphase cooling flow component in the
spectra. Previous \asca\ studies were divided between indicating a
massive cooling flow \citep[$\ga 60\msunyr$,][]{baue00b} and no
cooling flow \citep[$\la 40\msunyr$,][]{daw00}.

Using a standard model of a gas cooling at constant pressure as
implemented in our previous studies \citep[e.g.,][]{buot03a}, our
\chandra\ results indicate a best-fitting cooling rate of zero with
90\% upper limits of $\approx 5$~\msunyr\ in all annuli except annulus
1 (referring to the centroid case), where $\mdot = 7 (<11)$~\msunyr\
(90\% confidence). (We note that sometimes evidence for a cooling flow
and multiphase gas in clusters has been claimed based on the analysis
of CCD spectra extracted from a large aperture covering, e.g., the
entire putative cooling flow region \cite[e.g.,][]{clar04a}. However,
the distribution of temperatures produced by the radial variation of a
single-phase medium within the putative cooling flow aperture, as well
as that contributed by the projection of higher temperature gas
outside the aperture if the data are not deprojected, generally cannot
be distinguished from a two-temperature medium or a cooling flow
\citep[e.g.,][]{buot99a,buot03a}. Hence, we have explored the need for
multiphase gas using the narrowest annuli allowed by the data
quality.)

Although the hardness ratio map displayed no significant azimuthal
variations (\S \ref{image}), we searched for azimuthal spectral
variation by dividing up the annuli into four equal sectors We
rebinned the radial bins by a factor of two to achieve enough signal
for interesting constraints. When fitting a single \apec\ component to
each sector we found no evidence for azimuthal spectral variations,
consistent with the findings from the hardness ratio map. The
temperatures and iron abundances in each sector within an annulus were
found to be consistent within their $\approx 90\%$ confidence
limits. In particular, we do not find the $>15$~keV temperature spike
in a region a few arcminutes to the East of the center reported by
\citet{baue00b} from \asca. (The orientations of our sectors were
oriented approximately to match those of sectors 2-5 of
\citealt{baue00b}.)

Finally, we have studied the sensitivity of the results to various
sources of systematic error which we summarize below:

{\it Deprojection:} We performed a non-parametric deprojection using
the ``onion-peeling'' method as implemented in our previous studies
\citep[e.g.,][]{buot00c,buot03a,lewi03a}. That is, one begins by
determining the spectral model in the bounding annulus and then works
inward by subtracting off the spectral contributions from the outer
annuli. This technique tends to introduce fluctuations between
parameters in adjacent radial bins.  We found that to obtain results
of comparable quality to the projected analysis it was necessary
either to fix the temperature at a constant value near 8~keV or
restrict the radial variation in the the iron abundance so that its
logarithmic slope was $\la \pm 1$. After imposing either of these
restrictions we obtain results consistent with those reported above
for the projected analysis -- both in terms of fit quality and
parameter values. (We note that these results for the temperatures and
iron abundances obtained using our implementation of the
``onion-peeling'' method are very consistent with those obtained using
the deprojection scheme provided in \xspec\ using the {\sc projct}
routine and require the same restrictions on the temperatures and iron
abundances noted above.)

{\it Galactic Column Density and Bandwidth:} When energies below $\sim
1$~keV are included in the fits then a model assuming the nominal
foreground Galactic column density ($6.4\times 10^{20}$~\cmsq)
overestimates the flux at low energies, especially near
0.5-0.6~keV. If \nh\ is allowed to vary the fits are improved
substantially within each annulus; e.g., in annulus 4
$\Delta\chi^2=45.9$ for 228 dof. The best-fitting column densities are
$\approx 14\times 10^{20}$~\cmsq -- about twice the Galactic
value. Since the excess column densities apparently do not vary with
radius it is likely that the excess is associated with an
underestimate of the Galactic column. (Using the background models
described below we verified that the inferred excess column density is
not the result of over-subtracting the soft background using the
standard templates of nominally blank fields.) Our result is similar
to that reported by an \xmm\ study of A478 \citep{poin04a}.

Compared to the results presented above for energies above 1.5~keV,
the temperature and iron abundance are mostly unaffected when fitting
down to 0.5~keV with a column density about twice the Galactic
value. The most significant change is a systematic decrease of
$\approx 0.7$~keV in the temperature values in all annuli, but the
shape of the radial temperature profile is preserved. Although we
expect the excess column density is of Galactic origin we preferred to
confine our analysis to energies above 1.5~keV to avoid sensitivity to
the precise value.

{\it Background:} As described in \S \ref{obs}, for comparison to
results obtained using the standard blank background fields we also
constructed a model background by fitting the total (source plus
background) spectrum of regions of the ACIS-I as far away from the
center of A644 as possible. We found that results obtained using this
modeled background agreed extremely well with those of the standard
blank fields. As expected, by far the largest differences were
observed in annulus 9; e.g., $T=8.6\pm 0.6$~keV using the modeled
background compared to $T=8.1\pm 0.5$~keV for the standard blank
fields.

{\it Plasma Code:} We investigated the sensitivity of our results to
the plasma code using the \mekal\ model. The quality of the fits and
the temperature values were found to be very consistent within the
$1\sigma$ errors. The iron abundance values also were consistent with
those obtained using the \apec\ code, though the values obtained with
the \mekal\ code were generally smaller by 10\%-20\%. These results
are consistent with previous studies \citep{buot03a,buot03b,hump04b}.

{\it Calibration:} We explored the sensitivity of our results to the
version of the \chandra\ calibration using previous versions of {\sc
ciao} (v3.0.2) and the {\sc caldb} (v2.26). We found the results on
the fitted spectral parameters to be consistent with those quoted
above within the estimated $\approx 1\sigma$ statistical errors.

\section{Gravitating Mass and Gas Fraction}
\label{mass}

\begin{figure*}[t]
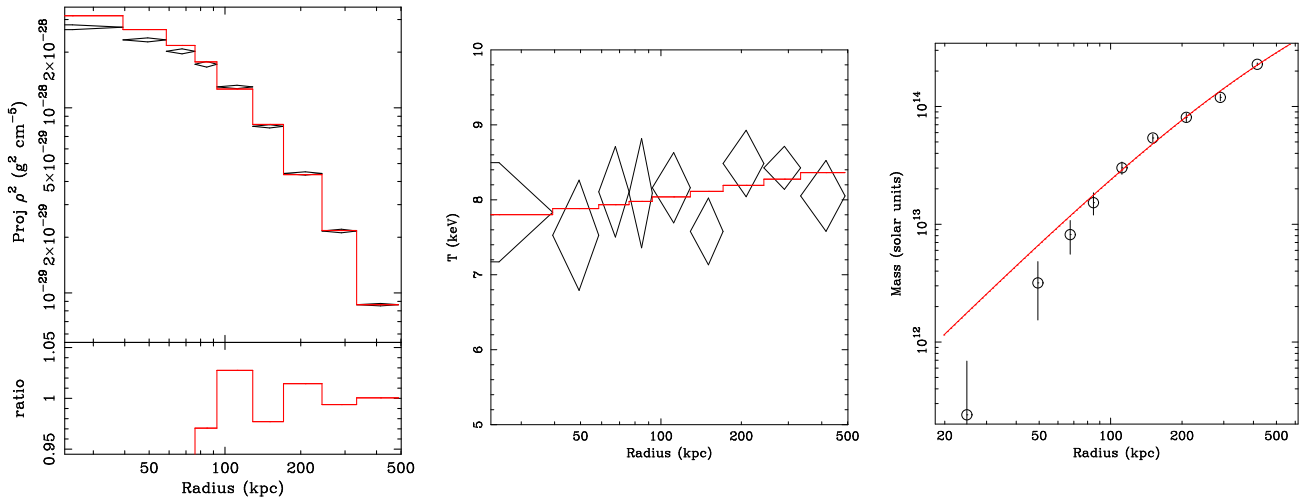

\parbox{0.32\textwidth}{
\centerline{\psfig{figure=f4a.eps,angle=-90,height=0.27\textheight}}}
\parbox{0.32\textwidth}{
\centerline{\psfig{figure=f4b.eps,angle=-90,height=0.23\textheight}}}
\parbox{0.32\textwidth}{
\centerline{\psfig{figure=f4c.eps,angle=-90,height=0.23\textheight}}}
\caption{\label{fig.profile} \footnotesize
Models fitted to radial profiles (centroid case) as discussed in \S
\ref{mass}. In each case the inner three data points are excluded from
the fits. ({\it Left}) \chandra\ radial profile of the projected gas
density squared ($\int \rhog^2 \Lambda(T,Z_{\rm Fe})dl / \Lambda_{\rm
ann}$) obtained by dividing the $norm$ parameter of the \apec\ model
(Table \ref{tab.spec}) by the area of the annulus; $\Lambda_{\rm ann}$
is the plasma emissivity evaluated using $T$ and $Z_{\rm Fe}$ of the
annulus.  The diamonds represent the binned data and the solid line
the best-fitting double-beta model. The fit residuals are plotted in
the lower panel. ({\it Center}) Temperature profile of the binned data
(diamonds) and best-fitting power-law model. ({\it Right}) Mass
profile (circles and error bars) and best-fitting NFW model.}
\vspace{-0.5cm}
\end{figure*}

\begin{table*}[t] \footnotesize
\begin{center}
\caption{Deprojected Radial Profiles of Gas Density and Temperature}
\label{tab.gastemp}
\begin{tabular}{lcccccccc} \tableline\tableline\\[-7pt]
\\[-1pt]
                                                                                                                                                        
\tableline\\[-5pt]
\multicolumn{8}{c}{Gas Density}\\[+2pt]
\tableline
\\[-7pt]
                & & $r_{c}$ & & $\rhog_0$ & $r_{c,2}$ & & $\rhog_{0,2}$ \\
		Model & $(\chi^2$/dof) & (kpc) & $\beta$ & ($10^{-26}$ g cm$^{-3}$)
		& (kpc) & $\beta_2$ & ($10^{-26}$ g cm$^{-3}$)\\

\\[-7pt]
\tableline\\[-5pt]
$1\beta$         & 12.1/3  & $120\pm 8$ & $0.64\pm 0.02$  & $2.27\pm 0.12$\\
$2\beta$         &  4.3/0  & $213\pm 33$ & $2.0\pm 0.3$  & $2.19\pm 0.24$
			   & $352\pm 115$ & $1.1\pm 0.3$  & $0.82\pm 0.17$\\
\tableline\\[-7pt]
\\[-1pt]
\\
                                                                                                                                                        
\tableline\\[-5pt]
\multicolumn{8}{c}{Temperature}\\[+2pt]
\tableline
\\[-7pt]
                &                & $T_{100}$    &         \\
Model           & $(\chi^2$/dof) & (keV)        & $p$     \\
\tableline
power law       & 2.7/4         & $7.9\pm0.4$ & $0.04\pm 0.05$ \\
                                                                                                                                                        
\tableline \\
\end{tabular}
\end{center}
\tablecomments{These models were (emission-weighted) projected along
the line of sight and fitted to the data about the centroid excluding
the inner three annuli. $T_{100}$ is the temperature evaluated at
$r=100$~kpc and $p$ is the exponent. Note that the best-fitting value
of $\beta$ for the $2\beta$ model was not well constrained and pegged
at the upper limit we allowed in the fits.}
\end{table*}

\begin{table*}[t] \footnotesize
\begin{center}
\caption{Parameters of NFW Fit to the Gravitating Matter Profile}
\label{tab.mass}
\begin{tabular}{cccccccc}  \tableline\tableline\\[-7pt]
& Best & $\Delta$Statistical & $\Delta$Background & $\Delta$Plasma & $\Delta$Bandwidth 
& $\Delta$Calibration & $\Delta\rhog$ \\
\tableline \\[-7pt]
$c$ & 6.1 & $\pm 1.2$ & $-0.4$ & $+1.2$ & $-0.4$ & $+0.1$ & $+0.4$\\
\rvir (Mpc) & 2.9 & $\pm 0.4$ & $+0.3$ & $-0.3$ & $-0.2$ & $-0.1$ & $-0.1$\\
\tableline \\[-7pt]
\tableline \\
\end{tabular}
\tablecomments{The virial radius (\rvir) is defined to be the radius
where the enclosed average mass density equals $103\rho_c(z)$
appropriate for the $\Lambda$CDM concordance cosmology
\citep[e.g.,][]{eke98a}. The inner three annuli are excluded from the
fits. The ``Best'' column indicates the best-fitting value and
``$\Delta$Statistical'' the $1\sigma$ statistical error. The other
columns refer to systematic differences in the best-fitting values
obtained for different choices (see end of
\S \ref{spec}) in the background, plasma code, bandwidth, and
calibration. For the bandwidth case the data are fitted down to
0.5~keV using a fixed foreground column density of $14\times
10^{20}$~\cmsq (see end of \S \ref{spec}). The final column refers to
differences obtained when using a $1\beta$ model to parameterize the
gas density.}
\end{center}
\end{table*}

To determine the gravitating mass distribution from X-ray observations
requires the hot gas to be in hydrostatic equilibrium. Both N-Body
simulations and gravitational lensing studies confirm the reliability
of X-ray mass measurements, particularly for clusters with regular
X-ray image morphologies \citep[e.g., for a review see][]{buot03d}.
The assumption of hydrostatic equilibrium should be valid outside the
central $\sim 75$~kpc region of A644. Outside the core region the
derived radial variations in the density and temperature of the hot
gas agree for both the cases where the cluster center is defined as
the surface brightness peak or centroid.  Consequently, we exclude the
central three annuli (defined in Table \ref{tab.spec}) from our
default analysis, though for comparison we summarize results obtained
using all annuli.

The approach we use to calculate the gravitating mass distribution
follows closely our previous studies \citep{lewi03a,buot04a}. As is
standard we assume spherical symmetry to provide a spherically
averaged mass profile appropriate for comparison to other observations
and to the spherically averaged mass profiles obtained from
cosmological simulations. In the equation of hydrostatic equilibrium
we evaluate the derivatives of the three-dimensional gas density
(\rhog) and temperature ($T$) using simple parameterized models as
discussed below.

For our default analysis we projected parameterized models of the
three-dimensional quantities, \rhog\ and $T$, and fitted these
projected models to the results obtained from our analysis of the data
projected on the sky (Table \ref{tab.spec}). In this manner we
obtained good constraints on the three-dimensional radial profiles of
\rhog\ and $T$. Although our analysis closely follows our previous
studies, we note two minor improvements. First, when fitting models to
the radial profiles of the gas density and temperature to the data
binned in annuli on the sky (as given in Table \ref{tab.spec}) we
integrated the models over each radial bin (rather than only
evaluating at a single point within the bin) to provide a consistent
comparison. Second, when projecting models of the gas density and
temperature along the line of sight we also included the radial
variation in the plasma emissivity ($\Lambda(T,Z_{\rm Fe})$).

In Figure \ref{fig.profile} we show the radial profile of the
emission-weighted projection of $\rhog^2$; i.e., the ``norm''
parameter (see caption to Table \ref{tab.spec}) of the \apec\ model
divided by the area of the annulus. A single $\beta$ model provides a
good description of the data with fit residuals $\le 5\%$ (Table
\ref{tab.gastemp}). Because the inner three annuli were excluded, a
marginally larger value of $\beta$ was obtained compared to the
surface brightness fits (\S \ref{image}).  Adding a second
$\beta$-model component provides small improvement in the fits with
smaller residuals ($\le 3\%$) than the $1\beta$ model (Table
\ref{tab.gastemp}); the $2\beta$ model is displayed in Figure
\ref{fig.profile}.  Although the parameters for the $2\beta$ model are
not as well constrained as those for the $1\beta$ model we adopt the
$2\beta$ model for our analysis of the mass distribution because it
describes the data a little better. Note that if the inner three
annuli are included then the parameters obtained are consistent with
those in Table \ref{tab.gastemp} but are somewhat better
constrained. For the case where the annuli are oriented about the
emission peak the second component of a $2\beta$ model (fitted to all
annuli) matches very well the $1\beta$-model results obtained for the
centroided case listed in Table \ref{tab.gastemp}, demonstrating the
similarity of the gas density profiles obtained outside the core
region for the peak and centroid case.

The radial profile of the emission-weighted projection of $T$ is shown
in Figure \ref{fig.profile} (middle panel) along with the best-fitting
power-law fit. The simple power law is an excellent fit (Table
\ref{tab.gastemp}) and demonstrates that the temperature profile is
consistent with isothermal within the estimated $1\sigma$
errors. Notice that despite the inner three annuli being excluded from
the fits the model describes those data points well.

We constructed the mass profile using the models just described for
the gas density and temperature. Following previous studies
\citep[e.g.,][]{lewi03a}, for each annulus listed in Table
\ref{tab.gastemp} we assign a single (three-dimensional) radius value,
$r \equiv [(R_{\rm out}^{3/2} + R_{\rm in}^{3/2})/2]^{2/3}$, where
$R_{\rm in}$ and $R_{\rm out}$ are respectively the inner and outer
radii of the (two-dimensional) annulus. The radial mass profile is
plotted in Figure \ref{fig.profile} using the $2\beta$-model for
\rhog\ and the power law model for $T$.  (We emphasize that although
the mass values for the inner three data points are shown, being
derived from the gas density and temperature models, they are not
included in the fits.)

The NFW model is a good smooth fit when excluding the inner three
radial bins (Figure \ref{fig.profile}). We obtain $\chi^2=6.9$ for 4
dof with fit residuals $\le 15\%$. Because the mass data points are
correlated it is necessary to recalibrate the $\chi^2$ statistic to
examine goodness of fit using our Monte Carlo simulations (\S
\ref{spec}); i.e., for each set of parameters for the gas density and
temperature profiles obtained for each simulation, a mass profile is
constructed and then fitted with an NFW model. These simulations
indicate the the $\chi^2$ value quoted above is within one standard
deviation of the expected value.

In Table \ref{tab.mass} we give the parameters of the NFW fit and the
estimated statistical and systematic errors. The estimates for the
systematic errors are mostly based on our discussion in \S
\ref{spec}, and we have also included the differences obtained when
using a $1\beta$ model to parameterize the \rhog\ profile. (Note we do
not provide an estimated error arising from the deprojection method
because the statistical errors are considerably larger when using the
onion-peeling method.) The derived concentration parameter ($\approx
6$) and virial radius ($\approx 2.9$~Mpc) imply a virial mass of
$\approx 1.5\times 10^{15}\msun$ appropriate for a massive galaxy
cluster. The value of $c$ is consistent with that expected from CDM
\citep[e.g.,][]{thom01a,tasi04a}.

We mention that within the central region (i.e., the central three
annuli), where the assumption of hydrostatic equilibrium is suspect,
the mass density profile is flatter than NFW; i.e., the cumulative
mass profile at small radii lies below NFW as seen in Figure
\ref{fig.profile}.  However, for the case where the annuli are
centered about the emission peak we find the mass profile within the
central $\sim 70$~kpc is highly uncertain but consistent with NFW. (To
obtain a reasonable smooth fit to the centrally peaked temperature
profile in this case, we modeled the profile with two power laws with
exponential cut-offs at both ends.) The smaller mass values in the
core for the centroid case are attributed to the flatter density and
temperature profiles obtained with respect to the peak case; i.e.,
when the annuli are not positioned about the emission peak radial
differences in spectral quantities are smeared out.

\section{Conclusions}
\label{conc}

Outside the core ($R\sim 75$~kpc $\sim 0.03\rvir$) the hot ICM has
properties consistent with a (relaxed) cool-core cluster out to the
largest radii investigated ($R\sim 415$~kpc $\sim 0.14\rvir$). The
X-ray surface brightness is very regular, moderately elliptical in
shape, with no evidence of substructure; previous
\rosat\ studies generally indicated a relaxed cluster on such scales
\citep[e.g.,][]{buot96b}.  The radial profile of the density of the
ICM is fitted well with a single $\beta$ model, with a double $\beta$
model providing slight improvement. The temperature profile is
consistent with isothermal or gently rising
\citep[e.g.,][]{alle01c,vikh04a}. The entropy factor ($S\equiv
k_BT/n_e^{2/3}$, e.g., $S=797\pm 21$~keV~cm$^2$ at $R=0.14\rvir$) is
consistent with those observed in cool-core clusters
\citep[e.g.,][]{piff04a}. The iron abundance declines with increasing
radius \citep[e.g.,][]{degr04a}. The gravitating mass profile is well
fitted by an NFW profile with reasonable concentration parameter for a
halo of its virial mass. Finally, the gas fraction ($0.076\pm 0.002$
at $R=0.14\rvir$) is quite typical of relaxed clusters observed by
\chandra\ and produced in CDM simulations \citep[e.g.,][]{alle04a}.

Inside the core, however, the \chandra\ data reveal an irregular,
disturbed system. The X-ray image displays asymmetrical isophotes
about the emission peak which are offset from the global emission
centroid by $\approx 50$~kpc. When oriented about the emission peak,
the temperature profile of the ICM declines towards the center (down
to $R\sim 50$~kpc), very much resembling a cool-core cluster. But at
small radii the temperature profile turns around and increases toward
the center; i.e., the inner temperature profile is inconsistent with a
constant at the $2.3\sigma$ level. Both the iron abundances and
entropy profiles mirror the shape of the temperature profile with an
excess at the center.

Since A644 does not possess significant radio emission
\citep{burn90a}, and the \chandra\ image does not show any X-ray
cavities, the suppression of a cooling flow cannot be attributed to
the present day feedback from an AGN.  Nor is it easily ascribed to
heating by a past radio outburst from the black hole in the cD,
because the synchrotron lifetime of 1~GHz-emitting cosmic rays is
$\sim 10^8$~yr for a magnetic field strength of a few $\mu$Gauss in
the cluster ICM. Since this energy loss timescale is comparable to the
sound-crossing time for the cluster core (and thus to the time needed
to heat the core), the core should remain hot for about the same time
as the 1~GHz radio emission remains visible after heating is
terminated. It is even more difficult to explain heating of the ``hot
spot'' (i.e., the central "hot" radial bin oriented about the emission
peak -- see Figure 2) by an AGN in the recent past, because an even
shorter time since the cessation of cosmic ray injection is required
due to its smaller physical size.

Nevertheless, a search for a steep-spectrum radio source would be
worthwhile. The absence of low-frequency radio emission \citep[e.g.,
VLA Low-frequency Sky Survey;][]{cohe04a} would be an even more
powerful test against the AGN heating model in A644, because
100~MHz-emitting cosmic rays have three times longer lifetimes in
comparable magnetic fields.  Moreover, although the higher temperature
in the ``hot spot'' appears to be accompanied by a sharp increase in
iron abundance, supernovae are a most unlikely heating source for even
this small portion of the core because the required number of
supernovae would need to be accompanied by a recent, galaxy-sized
($\ge10^9\msun$) starburst. Neither a blue galaxy nor a luminous
emission-line nebula are visible in the cluster core of A644
\citep{hu85a}.

However, the morphological irregularities in the core of the X-ray
image are very likely the result of merging. The cD galaxy (2MASX
J08172559 -0730455) is located $\approx 20$~kpc South of the X-ray
emission peak nearly along the direction of the centroid shift noted
in \S \ref{image}. It is also $\approx 40$~kpc displaced from the
centroid of the X-ray emission computed from larger radii. Evidently
the cD galaxy and core ICM are ``sloshing'' in the potential well of
the main cluster, and this energy has prevented the establishment of a
cooling flow.  The metallicity peak measured within the innermost
radial bin probably indicates that previous merging did not destroy
the initial peak, created when it was spatially coincident with the cD
galaxy. This is because, as argued above, the peak is unlikely the
result of very recent supernova enrichment. Also, the visual
appearance of the A644 galaxies is quite diffuse, stretching $\sim$ 2
Mpc linearly across the sky, suggesting a system far from
relaxation. This morphology strengthens the argument for an on-going
merger.

The properties of A644 (this paper), A2029 \citep{lewi02a,lewi03a},
and A2589 \citep{buot04a} have important similarities. Their X-ray
images indicate a peak which is offset from the centroid of the global
X-ray halo, with A644 displaying the most extreme shift ($\approx
60$~kpc offset) compared to the others (A2589: $\approx 10$~kpc,
A2029: $\approx 4$~kpc). Other than these offsets, the X-ray emission
is very regular in each case with no indications of cavities as are
seen in cool-core clusters with AGN. \footnote{We remark that
\citet{clar04a} have emphasized the existence of low-level ($\approx
10\%$) surface brightness deviations from a smooth elliptical model in
the central $\sim 10$~kpc of A2029 using the AO-1 \chandra\ image, and
claim that the WAT morphology anitcorrelates with these small
fluctuations. Using a more recent, deeper (80~ks) Chandra exposure we
confirm $\approx 10\%$ deviations in the surface brightness
(Zappacosta et al.\ 2005, in prep), but we do not confirm a strong
anticorrelation with the WAT morphology as seen in other cool-core
clusters. Compared to other cool-core clusters with obvious central
disturbances these $\approx 10\%$ deviations in the surface
brightness, which translate to even smaller ($\approx 5\%$) deviations
in the gas density, indicate that the core of A2029 is in a much more
relaxed state. For comparison, we find that Perseus, a cool-core
cluster with an obvious distrurbance in the core, displays up to 50\%
fluctuations in the surface brightness. It should be added that, unlike
A2029, Perseus also has strong azimuthal fluctuations in the
temperature and metallicity which, however, apparently consipire to
produce approximate azimuthal balance in the gas pressure and
therefore approximate hydrostatic equilibrium \citep{sand04a}.}  Two
of these clusters have no detected AGN radio emission, while A2029 has
a WAT morphology AGN, very unusual (unique to our knowledge) for a
cool-core cluster. Finally, the lack of nebular emission lines in the
cD galaxies are characteristic of non-cool-core clusters
\citep{card98a}.

Of these three clusters only A644 possesses a central temperature
peak. Although A644 appears to be unique in that its temperature
profile declines toward the center and then rises, there are a few
reports of clusters in the literature with temperature profiles
measured by either \chandra\ or \xmm\ that peak at their centers. The
clusters A2218 \citep{govo04a,prat05a} and A3921 \citep{bels05a} are
violent, advanced mergers with with pronounced asymmetrical surface
brightness and temperature distributions uncharacteristic of A644,
A2029, and A2589. Another cluster with a reported central temperature
peak is A401 \citep{sake04a}, which is in the early stages of a major
merger (with A399). Nevertheless, \citet{sake04a} argue that the
properties of A401 are determined primarily from its own hierarchical
formation before interaction with A399. In this context it is
interesting that A401 has a weak radio halo since we have suggested
that radio halos form only in massive clusters where a violent merger
has proceeded fully into the cluster core \citep{buot01b}. If the
other clusters had a similar history, their lack of radio halos
indicate they are presently at much more evolved states than A401.

In the paradigm wherein the energy output from an AGN disrupts cooling
flows, it is expected that just prior to the onset of AGN feedback a
cluster will have the characteristics of a cool-core cluster but
without significant radio emission or X-ray cavities. Indeed, the
clusters A644, A2029, and A2589 are excellent candidates for this
pre-feedback state, with A644 being at a relatively early state and
A2029 at the very latest evolutionary state, where an AGN outburst has
already occurred but little or no heating from AGN-generated cosmic
rays has commenced. The extremely advanced evolutionary states of
A2029 and A2589 make them ideal for studies of their core mass
profiles using the assumption of hydrostatic equilibrium.

%\vskip -1cm

\acknowledgements 

We thank A.\ Lewis for his participation in this project at early
stages. J.T.S.\ thanks A.\ Venketasen for useful discussions of
cluster heating mechanisms. D.A.B.\ acknowledges partial support from
NASA under grant NAG5-13059 issued through the Office of Space Science
Astrophysics Data Program.  Partial support for this work was also
provided by the National Aeronautics and Space Administration through
Chandra Award Number GO1-2130X issued by the Chandra X-ray Observatory
Center, which is operated by the Smithsonian Astrophysical Observatory
for and on behalf of the National Aeronautics and Space Administration
under contract NAS8-03060.

%XXX bibtex bibliography \\
%\bibliographystyle{apj}
%\bibliographystyle{apj_hyper}
%\bibliography{dabrefs}

\end{document}